\documentclass[epjST]{svjour}
\usepackage{graphics}
\usepackage{color}
\usepackage{hyperref}
\usepackage[english]{babel}
\usepackage{mdframed,xcolor}
\hypersetup{
  colorlinks=true,
citecolor=blue,
linkcolor=blue,
filecolor=blue,
  urlcolor = blue}

\begin{document}
\title{Complexity Aided Design:}
\subtitle{the FuturICT Technological Innovation Paradigm}
\author{Anna Carbone\inst{1,2}\fnmsep\thanks{\email{anna.carbone@polito.it}}  \and Marco Ajmone-Marsan\inst{1,3} \and Kay Axhausen \inst{2} \and Michael Batty \inst{4}\and Marcelo Masera \inst{5}  \and Erich Rome \inst{6}}

\institute{Politecnico di Torino, Italy  \and ETH Zurich, Switzerland \and IMDEA Network, Madrid, Spain  \and UCL, UK, \and JRC, Petten,  The Netherlands \and  Fraunhofer IAIS, Germany}

\abstract{``In the next century, planet earth will don an electronic skin. It will use the Internet as a scaffold to support and transmit its sensations. This skin is already being stitched together. It consists of millions of embedded electronic measuring devices: thermostats, pressure gauges, pollution detectors, cameras, microphones, glucose sensors, EKGs, electroencephalographs. These will probe and monitor cities and endangered species, the atmosphere, our ships, highways and fleets of trucks, our conversations, our bodies--even our dreams ....What will the earth's new skin permit us to feel? How will we use its surges of sensation? For several years--maybe for a decade--there will be no central nervous system to manage this vast signaling network. Certainly there will be no central intelligence...some qualities of self-awareness will emerge once the Net is sensually enhanced. Sensuality is only one force pushing the Net toward intelligence''.\\
These statements are quoted by an interview by Cherry Murray, Dean of the Harvard School of Engineering and Applied Sciences and Professor of Physics.  It is interesting to outline the timeliness and highly predicting power of these statements.  In particular, we would like to point to  the relevance  of the question  ``What  will  the earth's new skin permit us to feel?"  to the work we are going to discuss in this paper. There are many additional compelling questions, as for example:
``How can the electronic earth's skin be made more resilient?''; ``How can the earth's
electronic skin be improved to better satisfy the need of our society?'';``What can the science of complex systems
contribute to this endeavour?''
} %end of abstract
\maketitle
\clearpage
\tableofcontents

\section{Introduction}
\label{intro}

Millions of networked devices, sensing, filtering, interpreting and transferring data, are being embedded in the physical world at a tireless pace \cite{Chui}. Massive amount of information bytes  about
online game players, customers, suppliers, and operators are captured at a constantly
increasing rate with this exponential growth continuously fueled via smart phones and social networks. This large pool of data--captured,
communicated, aggregated, stored and analyzed on the
fly-- become more and more essential to every
sector and function of the global economy \cite{Mckinsey,Benkler,Codata}. In this scenario, highly
innovative concepts and approaches need to be deployed for the effective governance
and full exploitation of the information, its collection and distribution, in the
interest of an increased knowledge, the ordered management and development
of our planet, as envisioned within the FuturICT endeavour \cite{N1}. Like other essential
factors of production such as hard assets and human capital, much of modern economic activity, innovation, and growth simply couldn't take place without data and
networked infrastructures. What do these phenomena imply? Is the proliferation of
this networked structure simply evidence of an increasingly intrusive and uncontrollable
world? How can Big Data play a useful role for society's well-being? To build upon and capture value
from these processes and Big Data, academic institutions, public and private research centers are urged to develop
and deploy:
\begin{itemize}
  \item Efficient Technologies for sensing, storing, computing;
  \item Complex Systems Methodologies for analysing and extracting meaning from data;
  \item Suitable Policies and Protocols for the governance of data and processes.
\end{itemize}

The range of technology challenges will broadly differ,  depending on the maturity of the application areas. Legacy systems and
incompatible standards  too often prevent the integration of systems
and data on which the more sophisticated analytics should be built to create value from big data. New
problems and growing computing power will spur the development of new analytical techniques.
\par
 Our society is unable to understand and cope with the complexity of
today's techno-socio-economic systems. There is a
need for ongoing innovation in technologies and techniques to support individuals
and organizations to integrate, analyze, visualize and consume the growing torrent
of information generated by our connected world. The present effort deals with those technologies that are making possible
and functioning the network of networks, which are nowadays a main asset of our economy. It is obvious
that writing a paper covering all the technological challenges related to
the networked systems where we live would be highly ambitious and probably impossible to accomplish. We limit ourselves to a few remarkable
examples trying to reason on why these represent challenges for our economy and
society. We call our endeavour ``Technology Exploratory", which is an element of the
more complex structure  of Exploratories including the ``Society Exploratory" \cite{N7},
``Economy Exploratory"  \cite{N8_19} and ``Environment Exploratory"  \cite{N10} envisaged as a networked Research Infrastructure, within the overall FuturICT structure
(as sketched in Fig. 3 of  \cite{N1}). This integrated set of ``Exploratories" would ultimately become an
instrument for enabling an improved understanding, creating a basis for policy making based on a cross-disciplinary approach to the system of networks. The set
of Exploratories will also operate for capacity building by:

\begin{itemize}
\item	 supporting experts in the key FuturICT disciplines;
\item	 attracting and training young researchers in a multi-disciplinary environment;
\item	 boosting the number of experiments in system-of-systems investigations/explorations
\end{itemize}

The Exploratory concept revolves around the FuturICT vision \cite{N1} that it is essential
to exploit the power of information and communications technologies, to engage
researchers, policy-makers and a range of public interest groups to inform, alleviate
and solve critical problems, often referred to as `grand challenges', dealing with energy, aging, migration, climate change, competition, and social equity. Due to the strong
connections across techno-socio-systems, all these problems require powerful models
and tools in their resolution. In particular, new approaches to complexity --focused on
ways in which scientists and policy makers can come together to provide informed
solutions to these questions-- are central to the models that will be pioneered, extended and applied in the Exploratories. New approaches to computation and new ways
of organising ICT can spearhead new ideas to the governance and design of future
societies, and provide new capabilities to the interaction among all actors and technical systems and devices. In this framework, smart technologies can be used to underpin
city planning, transportation management, energy provision and usage, security and
privacy all in the wider context of the grand challenges that currently pervade our
society.

\subsection{Scientific and Societal Challenges}

The Technology Exploratory will respond in concert with the other FuturICT components to the key grand challenges with respect to sensing, measuring, and mining
physical and social data, within the wider FuturICT platform context to ensure that our
understanding of these challenges and advances is coordinated and integrated \cite{N2_6,N3,N4_5}.
This is a prerequisite that reflects the way ICT can be used to inform and solve socio-
economic problems, formulate and explore future policies, to implement the output
of the investigation to our social condition. A few major directions can be identified
within this perspective:

\begin{itemize}
\item Integrating and coordinating ``Systems of Systems" information concerning everyday citizen life. Information and Communication Technologies are being developed to increase the efficiency of energy
systems, to improve the delivery of services ranging from utilities to retailing in
cities, to improve communication and transportation, and to guarantee the security of these
new technologies. Integration and coordination of diverse systems across
functional and organizational boundaries, and even across jurisdictional and national borders, is crucial to ensure that value is added and that the efficiencies
realised are not dissipated.
  \item Understanding ``Systems of Systems" behaviour by integrated thinking as envis-
aged by Complexity Sciences . During the last 20 years, many new approaches
to modeling and simulating techno-socio systems have been devised that involve
holistic thinking. These approaches require further development in order to
make the policies concerning smart cities, transport and energy much more resilient
than hitherto. The great challenge is to foster convergence of interests and concerns of private and public actors involved. As ``Systems of Systems" have no
single master, the optimal solution satisfying all actors would rarely exist.
  \item Ensuring equity, fairness and achieving a better quality of life through new technologies. Efficient economic development must be balanced with equity, while new technologies have a tendency to polarise and divide at many levels. Efficient ways in
which new forms of regulation can be developed and implemented leading to
better quality of life for all  should be paved. Other solutions might foresee
the empowerment of citizens, both as individual and as members of social networks, in
their interconnections with techno-socio systems and infrastructures.
\item Ensuring widespread participation through new technologies. New social ICT is
essentially network based and enables everyone to interact with those systems
across many domains and scales. Part of the process of coordination and integration, by using state of the art data systems and distributed computing, must
involve ways in which the citizenry is able to participate and  blend personal
knowledge with that of experts who are developing these technologies. The rate of appearance and set of capabilities of those
technologies might overwhelm the end user, affecting social acceptance and social
engagement. Privacy concerns as well as security are key to this challenge.
\end{itemize}

As reported in \cite{WEF},
``as power shifts from the physical to the virtual world, a new
paradigm for ensuring a health digital space should emerge...Online security is a public good and new mechanisms are urgently required to secure private investment in
exploring existing system vulnerabilities before they can be exploited". In the same
report, five main category of risks are outlined: Economic, Environmental, Geopolitical, Societal and Technological. It is worth noting that critical system failure and
cyber attack are ranked first respectively according to the impact - or the likelihood-
these events might have on society. An overview of the interconnection between technological and social risks is reported in  \cite{AON}. In this virtual structure
of our society, a paradigm shift is required in defining new leadership models and
organizational scheme. Leadership and organizational models of the
last century are mostly based on top-down, bureaucratic paradigms suitably working
for an economic system based on ``physical production'' \cite{Uhl-Bien}.  The leadership should emerge as a result of adaptive outcomes -learning, innovation and adaptability- analogously to the
complex dynamics of interactions in the new virtual mode of
economic and social organization. It is important to
notice that  the advantages of an organizational structure based on an adaptive mechanism of self-organization was already argued  in \cite{Selznick}. Today these concepts have become urgent at larger scales and level due to the increasing complexity of multilevel and multiscale interactions pervading every sector of our current societal organization.
\begin{table}
\caption{A list of the 20$^{th}$ Century Innovations Challenges according to their importance for social well-being \cite{Constable}.}
\label{tab:1}
% Give a unique label
% For LaTeX tables use
\begin{tabular}{ll}
\hline\noalign{\smallskip}
1 & Electrification \\
2 & Automobile \\
3 & Airplane  \\
4 & Water supply and distribution \\
5 & Electronics\\
6 & Radio and Television\\
7 & Agricultural mechanization\\
8 & Computers \\
9 & Telephone \\
10 & Air conditioning/refrigeration \\
11 & Interstate Highways \\
12 & Space flight \\
13 & Internet \\
14 & Imaging \\
15 & Household appliances \\
16 & Health technologies \\
17 & Petrochemical technology \\
18 & Laser and fiber optics \\
19 & Nuclear technologies \\
20 & High-performance materials\\
\noalign{\smallskip}\hline
\end{tabular}
\end{table}

\subsection{Innovation Challenges}
Innovation plays a vital role in the development of new business concepts, processes
and products, driving growth and opportunity in new markets and breathing new
life into a mature industry. As businesses are gradually and globally expanding, innovation will become a leading industry differentiator. In the tough battle to match and  conquer
the customers dreams and desires, innovation is a prerequisite for success and survival. Public and private research centers
can rapidly lose reputation and market shares if they fail to translate their innovation into practise. Failure
to innovate and the risk of technology/system failure entered the Top 10 Rank List
for the first time in 2011 \cite{AON} since the start of \emph{Global Risk Survey} in 2007.
\par
The innovation challenges that we are going to face over the 21$^{st}$ century are expected to
emerge from the main technological innovations of the 20$^{th}$ century listed in Table 1
\cite{Constable} as the latter have followed up the industrial and infrastructural transformation
of the 19$^{th}$ century. The  20$^{th}$ century items are ranked in Table 1 according to the
benefits brought to people. One should note that most of those technological
achievements have been accomplished thanks to outstanding findings and breakthrough in fundamental
sciences. Their development and exploitation depended for the most part on the economic wealth they generated that ultimately contributed to make them even more universally available.
\par
Competitive strategies should not  be based only on the current
market conditions but also on where the industry is heading. The ``tried and true'' business model,
which proved successful in the past, can no longer be the only model to meet society
needs and ensure competitiveness. More than ever, innovation, speed and
adaptability are essential for competing in
the global economy. According to the ``World Intellectual Property Organization'', the
number of applications for global patents rose 10 percent in 2010, when the economic crisis had induced a significant drop in other activities. This number is expected to
grow steadily in the next two years. The 2011 survey highlights that the ability to
embrace and leverage technology is emerging as a dominant factor underlying many
of the key risks facing organizations. With the heavy reliance on their technological
infrastructure (especially ICT), businesses are becoming more vulnerable to system
failures, data breaches and social media exposure, causing business interruptions,
loss of customers and damage to reputation. The importance to develop appropriate measures to mitigate those risks will steadily increase as
private and public infrastructures rely more heavily on new ICT technologies. Critical factors to ensure good performance of any innovative business are:
\begin{itemize}
\item \emph{Continuity}, whose disruption
will lead to losses and may even result in bankruptcy.
\item \emph{Security of data}, which is vital to protect intellectual property
against espionage and other data against
theft or abuse.
\item \emph{Technological innovation and competition},
new technologies need to be identified that could offer
new  opportunities for businesses. The effects that
these technologies will have on politics and society and their indirect consequences need to be identified and managed.
\item \emph{Failure to attract or retain top talents}, Academies, private and public research centers and companies should innovate to create the most appropriate environment for attracting and retaining top talents and foster further innovation.
\end{itemize}

Key factors allowing the above conditions for innovation and development
are robustness of internal and external critical infrastructures (IT networks and telecommunication systems, energy and water supply, financial network system,
supply chain and transport) against natural disasters, cybercrimes and other relevant
hazards and threats.

\begin{figure}
\center
% Use the relevant command for your figure-insertion program
% to insert the figure file.
% For example, with the option graphics use
\resizebox{0.5\columnwidth}{!}{%
\includegraphics{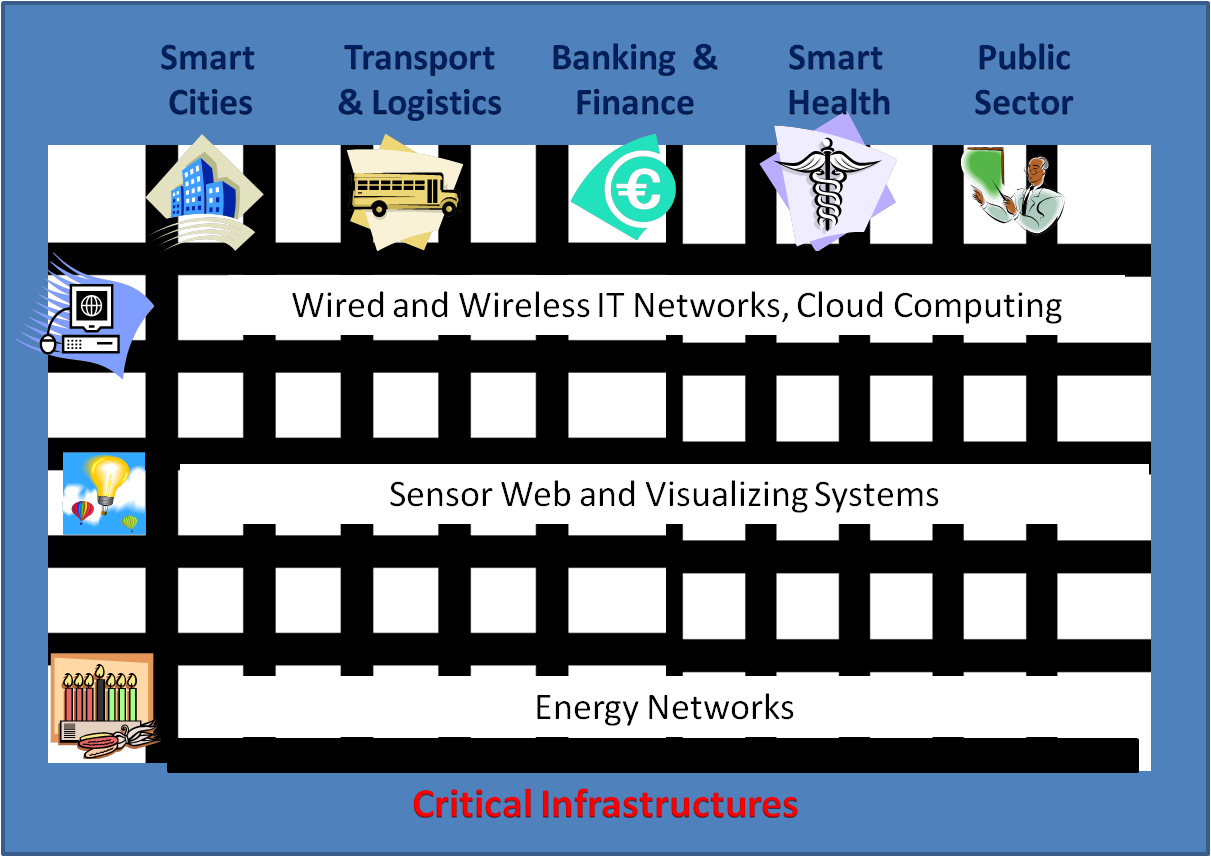}}
\caption{Scheme of the main components of the Technology Exploratory. The grid representation
allows to distinguish the horizontal cross-sectorial technologies and the vertical sectorial
technologies. The underlying frame corresponds to the issue of critical infrastructures which belong both to horizontal
and vertical technological sectors. All the components have close relationships to
the other Exploratories (Society, Economy, Environment) and to the main components
of the FuturICT Platform (Global Participatory Platform, Planetary Nervous System, Living Earth
Simulator).}
\label{fig:1}       % Give a unique label
\end{figure}
%
% For tables use
\begin{table}
\caption{Top 10 Risks: AON Global Risk Survey 2011}
\label{tab:1}       % Give a unique label
%% For LaTeX tables use
%\begin{tabularx}{0.75\textwidth}{llll}
%\begin{tabular*}{2cm}{llll}
\begin{tabular}{c p{3.55cm}p{3.55cm}p{3.55cm}}
\hline\noalign{\smallskip}
& 2011 & 2010 & 2009   \\
\noalign{\smallskip}\hline\noalign{\smallskip}
\footnotesize
%\baselineskip{1.5pt}
1 & Economic Slowdown & Economic slowdown & Damage to reputation\\[0.5ex]
2 & Regulatory  legislative changes & Regulatory/legislative changes & Business interruption\\[0.5ex]
3 & Increasing Competition  & Business interruption & Third-party liability\\[0.5ex]
4 & Damage to reputation brand & Increasing competition & Distribution or supply chain failure \\[0.5ex]
5 & Business Interruption & Commodity price risk  & Market environment\\[0.5ex]
6 & Failure to innovate or meet customer needs & Damage to reputation & Regulatory/legislative changes\\[0.5ex]
7 & Failure to attract or retain top talent & Cash flow/liquidity risk & Failure to attract or retain staff \\
8 & Commodity price risk & Distribution or supply chain failure & Market risk (financial)\\
9 & Technology or system failure & Third-party liability & Physical damage\\
10 & Cash flow liquidity risk & Failure to attract or retain top talent & Failure of disaster recovery plan \\
\noalign{\smallskip}\hline
\end{tabular}
\end{table}
\normalsize

\section{Scientific vision and State of the Art}
The FuturICT vision could not be conceived without our daily experience
of the Internet and wireless data networks, without our confidence
in  acquiring and transferring information about any human or artificial system and the awareness that the doors are now opened to creative ideas for
the exploitation of several new features. Telecommunication networks, and in particular wireless access systems, from mobile phone to WiFi, to newer technologies, such
as sensor networks, on one side, and the Internet on the other, have already transformed our world, leading to the society of connected individuals of the ``\emph{Internet
of People}".
\par
 It will soon take us to the next step: a world where also objects and machines are
interconnected among them, and with people, that is the basic concept underlying
the upcoming ``\emph{Internet of Things}"  which refers to sensors and actuators
embedded in physical objects, connected by networks to computers \cite{Chui}.
\par
The central
role of networking in the FuturICT vision reflects the way we live and work. Today, social relations, economic processes and markets, transport systems, health care systems, etc., largely
depend on networking for increased efficiency, reliability, security, and safety.
Models of transportation, urban planning and energy system operations are under
continuous development and refinement. One reason is that the way these systems
operate are strongly dependent on the development and availability of
new ICT technologies through the different sectors of human activities.  Another reason relates to the  continuous modification of social regional, national and
transnational organization. The overall result is that as more and more aspects of human activities become
interconnected by the availability of new technologies, the degree of complexity of
this evolving system increases. Hence a set of scenarios need to be envisaged
through a cross-disciplinary integrated long-term evolutionary vision.
Our society can benefit from integrating data, software and related protocols for the
purposes of building better cities, transport and energy supply systems. However
one has to consider that these systems are composed of autonomous sub-systems,
each one owned and operated by self-governing entities.
\par
Concepts
and tools traditionally belonging to the complex system science are
increasingly adopted to tackle the complexity of such techno-socio systems. Probabilistic models
describe speed drop by modeling vehicular flow as an open non-equilibrium system
of driven or active particles with energy generation/dissipation and possibly phase
transition phenomena in the global traffic behavior.
Complex network tools  have become increasingly exploited for describing
critical behavior of electrical power grids. Analogously to the multi-agent models
for financial investors, drivers are modeled for example as heterogeneous agents with
the level of heterogeneity (e.g. in car-following behavior) quantified by trajectory
observations recorded by means of GPS and cellular phone. Herding behavior concepts
are applied to car drivers, energy and good consumers to investigate the effects of
leader followers on the same lane, on the same power grid or commodity chain supply \cite{N12,N13,N11}.
\par
However,  much
remains to be done, for the improvement of today systems and their integration in a
``Planetary Nervous System"  that is capable of managing and integrating more and more
 diverse  amount of information than what we handle today, so as
to be able to understand and control the dynamics governing the internal behavior
of each domain, and those that from one domain influence the others. Much remains to be
done for the definition of new generations of networking able to scale to
the dimensions required to allow a seamless transition to the FuturICT vision, with
the necessary resilience and dependability. It must not be forgotten
that large network systems are critical infrastructures in themselves, and thus need
to be monitored, protected, controlled, and secured. These aspects are a fundamental but preliminary step required to address further more specific technological challenges
of FuturICT. Without scalable, robust, safe, secure, efficient networking support, the
FuturICT concepts might not be thought of.
\par
The research challenges to achieve these goals are huge. The amount of data to be
collected, transferred, processed and analyzed for the implementation of the FuturICT vision is humongous. The complexity inherent in the control approaches is still to
be discovered. Networks include different layers: they are not only technological, but
 also social and economical. Anticipating the networks of the future requires
experimenting with technologies and observe their adoption by society with diverse
adaptation processes. This multi-disciplinarity further complicates the research. However, the benefits that can result from the insights are enormous, and can drastically
improve the life of future generations through
 components
of the FuturICT project as for example the Planetary Nervous System \cite{N3}, the Living Earth Simulator \cite{N2_6}, the Global
Participatory Platform \cite{N4_5}, smart applications for cities \cite{N16}, energy \cite{N17}, education \cite{N22}
 and innovation services \cite{N20}.
\bigskip
\definecolor{shadecolor}{rgb}{0.941,1,0.941}
\begin{mdframed}[backgroundcolor=shadecolor]
\small
\begin{center}
\textbf{BOX 1:
Relevant Initiatives and Projects}\end{center}\par
\begin{itemize}
\item Open Data \url{http://make.opendata.ch/index.en.php}
\item Cisco Smart Connected Communities \url{http://www.cisco.com/web/strategy/smart_connected_communities.html}
\item Oracle Smart Cities \url{www.oracle.com/us/industries/public-sector/
smart-cities.htm}
\item Ericsson \url{http://www.ericsson.com/res/
/thecompany/docs/publications/business-review/2011/issue1/delivering-smarter.
pdf}
\item IBM Smarter Planet \url{http://www-05.ibm.com/innovation/uk/smartercity/}
\item Multiagent Transport Simulation  \url{www.matsim.org}
\item Google maps \url{http://maps.google.com}
\item Nokia maps
\url{http://europe.nokia.com/support/product-support/maps-
support/compatibility-and-download}
\item  OECD (2008): Protection of Critical Infrastructure and the role of investment policies
relating to national security, May 2008.
\item  OpenMI Association, \url{http://www.openmi.org/}
\item  Open Geospatial Consortium (2008a). Open GIS Keyhole Markup Language (KML)
Version 2.2.0.
\item  Open Geospatial Consortium (2008b). OpenGIS City Geography Markup Language
(CityGML) Encoding Standard, Version 1.0.0.
\item  TNOs Critical Infrastructure Disruption Database, Version 217, 14 June. TNO (2009).
\normalsize
\end{itemize}
\end{mdframed}
\bigskip
\bigskip

\subsection{Networking: from beneath Earth to the Clouds}
Wired and wireless data networks play a key role in many areas of FuturICT and
thus are one of the pillar research activities in the framework of the fundamental
Technologies.
\noindent
Data Networks are:
\begin{itemize}
  \item main building blocks for the achievement of main FuturICT goals, from
smart grids to intelligent transport and logistics, by collecting vast amount of data and acquire the necessary information about the
many variables of the systems to be monitored and controlled.
  \item main enabling technology for the protection of critical infrastructures, social participation systems, smart sensing environments, e-health services, and environment monitoring and protection, implementing data processing through cloud computing approaches, identifying
appropriate mechanisms for the effective control of the systems being monitored.
  \item relevant example themselves of critical infrastructure and of an essential service
to our society.
\end{itemize}
\noindent
In summary, network technologies are a key enabler for the distributed simulation
of the living planet over large numbers of computers, possibly adopting the cloud
concept and for the implementation of many of the FuturICT visions, from smart
cities to smart grids, from intelligent transport and logistics  to the protection
of critical infrastructures, from social participation systems  to smart sensing
environments, from e-health services to environment monitoring and protection.
While data networks have experienced an impressive development over
the last 20 years, the present level of technology is not sufficient to be painlessly
adopted `as is' for the achievement of the FuturICT goals. Indeed, on the one hand,
the number of data collection points and the amount of information to be carried
and filtered requires significant advances in architectural and technological aspects,
and, on the other hand, some peculiar characteristic of the FuturICT systems pose
challenges that have never been explored before as for examples:
\begin{itemize}
  \item \emph{IP Addressing}. This is a major problem of the present networks mostly based on IP version 4. The transition to version 6 has been encouraged
many times with very little success. The myriads of network terminals foreseen by
FuturICT require new addressing approaches. It might be that even such version
6 turns out not to be adequate, so that new alternatives must be explored.
  \item \emph{Resilience and Fault}. Todays networks include resilience and fault tolerance approaches, but they are based on the assumption that most network
elements are interconnected most of the time. The only exception that has received significant attention in the technical literature is given by the so-called
Delay Tolerant Networks (DTN), where only subsets of the network elements are
interconnected for significant portions of time. The FuturICT context (at least for
some applications, like, for example, environmental monitoring) may be such that
most of the network elements are disconnected most of the time. The possibility of
effective networking in such case has yet to be proven, and adequate architectures
have to be studied .
\item \emph{Security and privacy} are key issues for FuturICT, at different levels: the
data to be collected and transported are confidential, the injection of wrong data in
the system may lead to incorrect conclusions, the access to data and to conclusions
may be restricted to authorized individuals or terminals. The scale of the FuturICT
networks and the amount of information involved require to redesign the current
security and privacy approaches implying a paradigm shift in this area.
\item \emph{Continuous real-time monitoring of extremely large numbers of variables}. The use of such variable is required as input for the ``Living Earth Simulator". The correlation intrinsic in the series of observations for one variable
is normally very high, so that intelligent information filtering and data clustering
before transmission should be reasonably performed. However the setting of the
parameters for this operation is quite delicate, and requires serious and appropriate investigation.
\item \emph{Distribution of information storages}. The efficient access to information within networks is an issue which has been tackled by Content Distribution Networks
(CDN) and Peer-to-Peer (P2P) systems. Content Centric Networking (CCN) also addresses the same issues. The huge amount of information required for the
implementation of the FuturICT vision calls for new specific approaches.
\item \emph{Redundancy and network dependability}. Role of data networks in FuturICT
is vital and thus network resilience and dependability are crucial for the success
of the project. However, the size of the network and the amount of data to be
stored are extreme, so that providing redundancy implies very high cost. A very
careful investigation of the most effective approaches to achieve the desired level
of network dependability is a must.
\end{itemize}

The cloud is an emerging ICT technology that would act as a further
key component to catalyze the FuturICT platform \cite{Economist,Sluijs,Lasica}. The basic idea of
cloud computing is that many software functions are moved and operated by the
cloud rather than on individual computer facilities and servers. It is one of the highest expression of ``vertical integration" that will trigger new inventions ranging from
the basic technologies to the communication and multimedia sectors. A summary of
the main features of cloud computing can be found in \cite{NIST}: on-demand self-service,
broad network access, resource pooling, rapid elasticity and measured service. Cloud computing applications are usually optimized
to provide a simpler, easy-to-use interface, reducing the learning effort
and offering increased communication capacity between various software
packages. Cloud
computing is often considered an environmentally-friendly ICT technology  as the information resources and software applications are centrally maintained and
managed, thus reducing financial and energetic cost according to the
use.
 \par
 Nonetheless, the series of articles ``The Cloud Factories'' recently appeared on the \emph{New York Times} on September 23$^{th}$ 2012 evidenced that not everything is always green in the cloud.
\begin{itemize}
\item ``Online companies typically run their facilities at maximum capacity around the clock, whatever the demand. As a result, data centers can waste 90 percent or more of the electricity they pull off the grid ''.
\item ``Energy efficiency varies widely from company to company. But at the request of The Times, the consulting firm McKinsey \& Company analyzed energy use by data centers and found that, on average, they were using only 6 percent to 12 percent of the electricity powering their servers to perform computations''.
\item ``To guard against a power failure, they further rely on banks of generators that emit diesel exhaust. The pollution from data centers has increasingly been cited by the authorities for violating clean air regulations''.
 \item ``Nationwide, data centers used about 76 billion kilowatt-hours in 2010, or roughly 2 percent of all electricity used in the country that year''.
\end{itemize}
\par
Security and privacy are major concerns for many users that transfer their data
from locally-maintained servers to cloud computing systems. Especially since the very
nature of virtualization means data and applications stored in a way that allows them
to be easily accessible online. For this reason, better security methods must be employed in order to keep valuable information protected from intrusion or theft. Improved practices therefore include higher degrees of password protection, additional levels of security at the hosting site, and other advanced computing security
measures designed to protect the company's sensitive financial and proprietary information.

%Segaran,
\subsection{The Sensor Web: World Wide Sensing and Imaging}
The dramatic progress in ICT and Internet-based applications has the main consequence that individuals and organizations are exposed to an ever-increasing stream
of data \cite{Segaran}. In such a digital environment, the burden caused by information overload
and processing is largely compensated for by the extraordinary potential which the
availability of plentiful distributed data unleash for the management of our complex
environment. Once data is collected, stored and processed, users need to be able to
access, evaluate and explore their data by using friendly tools. Visualization provides a creative way to address the issues  by offering  means to manipulate
and make sense of large amount of data and the embedded information. They also
yield new capabilities to amplify cognition.  This sector of the FuturICT Technology Exploratory will be responsible to increase knowledge, to innovate and explore
the possibilities offered by visualization tools and more generally virtual reality technologies. Three main application areas will be highlighted with the relevant roles that information visualization can play as a tool to:
\begin{itemize}
  \item \emph{monitor and supervise spatially distributed systems}, such as land
territories and urban regions. Focus is needed on the methods and tools to integrate and make sense of concurring spatio-temporal relevant phenomena, such as
those concerned with urban security, environmental risks, energy and transport
management, thus supporting decision making in complex environments.
  \item  \emph{navigate through large data-base}, in this case, visualization helps to achieve an interpretation about data exhibiting meaningful complex patterns or correlations. This is the realm of visual analytics a scientific
field which is being established at the crossroads between design and computing.
  \item \emph{facilitate communication} with the aim to bridge the technological challenges
to the social need and knowledge for the co-generation of information communities. The role of information visualization in providing contexts for general public
information and thus adding social value to it are tasks particularly relevant to
the Exploratory of Technology.
\end{itemize}

Visualization is indeed an essential tool to analyze data generated in diverse
fields. Experiment and numerical simulations performed on systems belonging to social sciences, climate modeling, biological tissue are some examples of large amounts
of data that exhibit complexity features (self similarity over a wide range of spatial
and temporal scales). To quantify and understand features
like formation of patterns, clustered structures and  provide a realistic representation of such structure, efficient and accurate analytical tools are required.
\par
The available data sets, e.g. those related to public health and environmental science, are however highly heterogeneous and the
complex multivariate relationships among variables are often unknown. Integration
of traditional cartographic methods with those from information visualisation can
provide researchers and analysts with a range of tools for visually as well as statistically and computationally exploring these relationships.
\par
A urgent need
to go far beyond just the capability to make sense of huge data through visualization will relate to the  development of a ``Causality Discovery Visualization Technology" as argued
 in \cite{Chen1,Chen2} to understand and quantify causal relationships and
thus the underlying complex dynamics of visual structures.
 Moreover, imagine if we
could also virtually \emph{touch}, \emph{hear} and \emph{smell} while navigating in a virtual environment.
All our senses will cooperatively be conveyed  to remotely understand our complex world -as we do usually in our normal life. Computer applications need to be designed for
quantifying complex systems characterized by self-similar heterogeneity across space.
In particular, three-dimensional or higher dimensional forms of random heterogeneous media
are particularly interesting in this endeavour. The local microstructure can be quantified as
a three-dimensional fractional Brownian field and the relevant parameters
defined in terms of spatio-temporal correlation measures  \cite{vanKrevelen,Carbone1,Carbone2}.
\par
Development of sophisticated virtual reality tools is extremely timely for e-governance.
Visualisation platforms integrate virtual reality, geo-web, and visual analytics to empower stakeholders and citizens  in choosing and evaluating the
impacts of policies for their cities or regions. The main aim of this platform is to
clarify and communicate complex policies easily. Governments and Agencies can use
the platform to develop and implement policies with public participation, actively
engaging the public in the policy-making process \cite{Bazzanella,Jiong,Zhi-jie,Jie}.
\par
Large scale real-time three-dimensional visual simulations rely on very high computation performance and rendering efficiency with  further implied issues regarding interoperability, costs, flexibility, accessibility, usability, and information privacy. Key technological problems
 range from database file
format, real-time rendering algorithms and texture atlases, model simplification and
dynamic paging, multidimensional database management and parallel computing, remote visible and infrared sensing, photogrammetry and geo-information encoding,
shape grammars, web 2.0 techniques and web-services. It is nowadays possible to
integrate such enhanced models into the information layers of the internet creating
what is known as the ``Sensor Web". These advances have provided the opportunity to create innovative Information Systems to be widely exploited in urban land management, as intuitive interface to urban planning information and visualization of future
scenarios.
Virtual reality models  are  envisaged to overlap local data infrastructures for
planning in areas as transportation, logistic, energy, climate, air quality, fire propagation, health and public safety studies.
Image sensors directly measure the depth of objects, which provide an ideal data set
for urban modelling.  The quantity of data, image noise, lighting conditions, occlusions, and scene complexity complicate acquisition and elaboration. Automatic
modelling requires automatic structure segmentation and reconstruction, which is
complicated by occlusions between buildings or vegetation against buildings. Fusing
ground and aerial image sensors, aerial active sensors, and 2D footprint data from
GIS or CAD data can generate more accurate and automatic urban models (hybrid
sensor systems).
3D building models can be integrated into spatial databases and geographic information systems (GIS) to support urban planning and analysis applications. Geo-data
types (raster data and vector data) can be integrated into the 3D city model by draping them over the digital terrain model, which is a basic tool in 3D Geovisualisation. Further methods for the integration of vector data exist, such as visualising point features as 3D symbols or
extruding polygons to 3D blocks. Vast digital data resources
include geospatial referencing, ranging from precise geographic
coordinates, to street addresses and codes (such as zip codes and drainage basin indices). GPS receivers, PDAs, and cell phones generate an increasing fraction of these data.

\begin{figure}
\center

\resizebox{0.8\columnwidth}{!}{%
\includegraphics{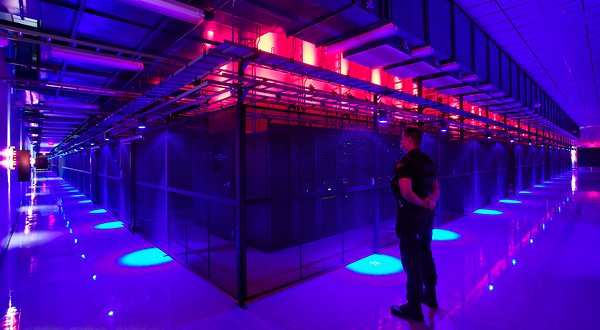}}
\caption{The Cloud Factories: Power, Pollution and the Internet. Data centers are filled with server, which are like bulked desktop computers, minus screen and keyboards, that contain chips to process data. (Photo Ethan Pines for \emph{The New York Times} 23 September 2012)}
\label{fig:3}       % Give a unique label
\end{figure}

\subsection{The Energy Web: World Wide Green Powering}

With the widespread availability of electronic control systems and data communication networks, and the emerging \emph{Internet of Things}, it becomes
possible to integrate in the electrical grid very large numbers of micro-generation
facilities. Micro-generation typically is not controlled by the grid operator, but by
homes or small offices which micro-manage production as well as end-user utilization,
possibly through the introduction of smart meters and time/load dependent tariffs.
This implies a leap in the electrical grid complexity, and leads to the concept of smart grid, whose objective is to increase the efficiency of electrical power production,
distribution and consumption. More precisely, smart grids are expected to be able to
accurately predict and intelligently respond to the behavior and actions of all players connected to the grid, from electric power suppliers, to consumers, in order to
efficiently deliver reliable, inexpensive, and sustainable innovative electricity services
\cite{N17,Qureshi}.
\par
Important issues to be approached over the next decades are: protection from security threats, including cyber attacks, integration of alternative power generation
sources with intermittent supply; reduction of peak demand surges; deployment of digital devices that can alter the nature of the electrical load problems.
\par
A truly smart grid concept will require technological advances at all levels. At the smallest scales increasing the effective sensitivity of infrared sensors , inventing new energy and
fuel storage systems (hydrogen), reducing or exploiting dissipation at the nanoscales, enhancing critical temperature in superconductors, turning spontaneous
fluctuation into energy . At the largest scale new approaches to data communications, automation, and distributed generation allowing greater capacity and
flexibility at lower cost.
The application of efficient digital processing and communications to the power grid,
allowing  on-the-fly collection and real-time redistribution of data about generation and consumption, so that entire grid can be controlled and optimized through
appropriate efficient algorithms.
\par
The necessary paradigm shift will imply  power consumption-follows-generation in future smart grids, as opposed to
traditional power grids  where power generation-follows-load. An example can be electric cars that can be
charged at night drawing on cheap power, and may be used during the day, and can
even provide access to the energy stored in their battery to absorb peak loads. More
in general, the following challenges should be targeted:
\begin{itemize}
  \item Adapting to variable conditions in supply and demand: a smart grid must be able
to effectively respond to events which occur anywhere in the power generation,
distribution and demand chain. Events can be of very different nature, from equipment failure, to clouds blocking the sun and reducing the amount of solar
power, to very hot temperatures requiring increased use of air conditioning, to
users leaving home and switching of most appliances.
  \item Creating innovative services: the widespread availability  digital technologies
will improve the efficiency, reliability and safety of power delivery at the
same time opening the way for entirely new services or improvements of the existing
ones.
  \item  Enhancing reliability, ensuring sufficient security and efficiency, providing self-healing: the real
time information collected from sensors and transmitted to distributed processing
elements that implement the smart grid control allow a smart grid to anticipate,
detect, and respond to system problems (including man-made attacks), and to
avoid or mitigate power outages, power quality problems, and service disruptions.
\end{itemize}

 GreenGrid  is one of the newly developed visualisation
techniques customized for power grid analytics built upon a collection of graph analytics and mining tools, that helps explore the planning and monitoring of the North
American power grids. The application of information visualisation holds tremendous
promise for the electric power industry, but its potential has so far not been sufficiently exploited by the visualisation community. Prior work on visualizing electric
power systems has been limited to depicting raw or processed information on top of
a geographic layout. Little effort has been devoted to visualizing the physics of the
power grids, which ultimately determines the condition and stability of the electricity
infrastructure \cite{Wong}.

\subsection{Smart Cities}
Many of the traditional functions of cities relating to movement and exchange that
enable agglomeration economies to be gained are being complemented and also replaced by digital networks and online behaviours which are not only making routine
functions more efficient but also enabling new qualities of life to be realised. The
development of routine ICT functions however are rarely coordinated and a plethora
of new networks now exist that supply new forms of information and enable different kinds of behaviour that do not so far enable real time economies to be gained.
In short, a massive breakthrough is required in connecting up such systems so that
value can be added and even more efficient routine functions enabled \cite{N16,Aurigi,Batty1,Batty2,Caragliu,Dutton,Harrison}.
\par
Parallel to this
is the deployment of new ICT for more strategic planning of the smart city. Indeed it
might be said that the cities will only become smart when routine ICT is assembled
so that a real guidance capability is developed for the city. In short, design, planning
and policy can and should be informed by the joining up of routine network systems
and the development of planning support and participation systems that deal with
longer term strategic vision of how cities can become more efficient and equitable
places. ICT has the potential to enable new governance systems for cities that build
on the idea of the Smart Cities. The following capabilities, methods and technologies
will be researched in the Technology Exploratory:
\begin{itemize}
\item Development of new methods of sensing movement in cities synthesising crowd
sourcing methods with more established methods of data collections from govern-
ment, marketing (geo-demographic), utilities and telecoms, as well as diverse social
media sources and sensor networks.
\item Development of new data systems with a city-wide focus in which data can
be merged, connected, and mined in the search for a new understanding of spatial
behaviour.
\item  Explorations of how new and largely invisible social and related contact networks
are enabling citizens to participate and learn about the state of their cities and
engage/participate in realising their vision through planning.
\item Examining the impact of new ICT on the form of the city, on land use activities
location and the design of new transportation networks, as well as on market
behaviour in the local economy.
\item Tracking of movements of people, materials and information that are rapidly
changing the quality of life in cities and provide the momentum for new competitive and innovative strategies for building ever smarter cities.
\item Identification of critical infrastructures and behaviours that generate risks and
crises that can be resolved by early warning systems which build on integrated
online data systems.
\item Exploration of booms and busts in urban cycles such as housing markets and
retailing activities that integrate the new understandings of complexity science
and relate more generally to other exploratories dealing with the financial system
and with political conflict in the nation state.
\item Development of new forms of urban governance fit for the information age that
build on existing and new online systems at all levels and across many spatial and
temporal scales.
\end{itemize}

\subsection{Transport and Logistics}
A broad range of theoretical developments to better understand traffic and transportation network performance, from the development of new theories to explain
traffic breakdown, car following, and traffic kinetics, to the development of new route
choice mechanisms, cooperative game behavior under network uncertainty, and dynamic models for travel activity generation \cite{Axhausen1,Axhausen2,Axhausen3,Helbing,HelbingM,HelbingN,Treiber,Neda}. The key issues are:
\begin{itemize}
  \item large scale integrated travel behaviour and logistics simulation models with open
time horizons and explicit learning mechanisms at the agent-levels as tools for
policy analysis.
  \item simplified representations of the travel demand and logistics systems for integrated
models across the infrastructure systems (energy, communications, water, long-
haul-logistics).
  \item improved network and schedule design optimisation for resilient operation.
  \item fast network design optimisation for the basic built infrastructures (roads, rail-
ways, canals, pipelines).
 \item data fusion and data aggregation processes to allow continuous system performance.
\end{itemize}
The collation of the available data streams of the European countries will require
their translation into a common terminology and where necessary adjustment to
make them comparable. The necessary contracts with their providers will have to
be negotiated to make their collection smooth and reliable (APIs instead of web-bots). Examples are traffic counts,
flight movements, container movements, shipping
rates, TCP/IP traffic, tourism
flows, mobile phone use, the US FAA 10 % ticket
sample, commercial air ticket data bases, EUROSTAT statistics, health warnings.
The Transport and Logistic Sector will derive the key leading indicators using the
best techniques to summarize the various data streams, while accounting for both
their temporal and spatial nature. The comprehensiveness of the indicators and the
underlying data streams will be continuously improved. We be filling the data gaps.
\par
Two data streams are missing to understand the dynamics of transport movements
in Europe: on the one hand, we are lacking the observations and the understanding of
the social network geography of the Europeans and how their movements and their
resources, and infectious diseases are channeled through them; on the other hand,
there is no easy point-of-access to costs of long-distance travel, as all operators use
price discrimination to maintain their margins.
\begin{itemize}
  \item  A set of representative social network surveys in a large sample
of European countries/regions, which cover the range of the current economic
development and history of migration is needed. The expectation is that further research
groups will join the initial effort over the years. FuturICT will make some support
for these teams available (data archiving; staff support; data analysis).
  \item The long distance travel market (air, rail, ferries, gasoline prices) is characterized
by strong price discrimination along the time of booking, season, service levels/class and origin/destination pair. We will set up an automated web-based
observatory which will sample the prices systematically across the continent for
European and Intercontinental journeys for a range of booking conditions (time to
departure, class, length of stay) employing a number of servers simulating different customers by location. The dataset is enriched with suitable data on booking
levels, school holidays, legal holidays, tourism
flows, etc. The data set will be
archived continuously and made available in regular intervals as a counterpart to
the official 10\% FAA ticket sample available for the US airline market.
  \item We will build the tools for crisis preparation in personal transport: The possible
crises are manifold: e.g. a volcano eruption disrupting air travel across Europe;
major population movements after a major chemical accident; shutdown of the
railway services during a developing epidemic. The authorities and firms need a
model describing the current situation and its underlying interactions and dependencies reliably with theoretically sound models. These tools will be the basis for
detailed models of disease spreading.
\item We will draw on the tools available for the GNU-public license framework at Matsim (see Box 1).
\item To support the tool we will model the impacts of the social network geographies
of the Europeans: Based on the new data collected the project will model the
interactions between the frequencies of contacts across all modes and the political,
economic and transport performance across Europe and where appropriate the
world.
\item We will build the tools for crisis preparation in logistics: The market for logistics
services is more complex than the market for passenger transport due to lengths
of the supply chains and the larger number of decision takers and actors involved.
Based on on-going work to expand agent based tools with the suitable data structures and models, we need to increase the computational speed to match the new
scale of the implementation.
\item We will implement the tools for Europe so that related projects in the FuturICT
and elsewhere can explore scenarios of interest to them. We will consolidate the
necessary data, choice models, generate the agent population, establish the net-
works, calibration and validation data for the first implementation and then the
necessary biannual updates and five-years major updates. The contacts with the
users and their experiences and results will be integrated on an on-going basis:
the living model will continuously adapt and learn!
\end{itemize}

Understanding the regional or big picture transportation status can be complex. Many
of today's issues involve new concepts that the public and even professionals in the
field have difficulty grasping. Examples of difficult-to-explain transportation problems
are: How does traffic build up when a blockage occurs on a freeway? Why does the
build-up take so long to dissipate after the blockage is gone? Why do ramp meters
work? How does congestion pricing play a similar role? How would a high-occupancy
toll lane work? How would it help nonusers? How do shrinking or expanding market sheds or economic corridors affect access to employment, employees, suppliers,
or a greater range of housing choices? Researchers combined OpenGL and modelling
techniques to develop an interactive virtual helicopter. The tool renders GIS (geo-
graphic information system) and transportation infrastructure data combined with
traffic sensor, transit, and accident data. The prototype system interacts with real-
time traffic databases to show animations of traffic, incidents and weather data. With
the virtual helicopter, users can travel through the region to inspect conditions from
different angles, at different distances. Users can monitor and interact with traffic
control devices such as dynamic message signs, closed-circuit TV feeds, and traffic
sensors. They can even track and view effects of ground or air vehicles equipped with
GPS transponders.
\par
The data from simulations at the petascale level are too large to store and study
directly with conventional post-processing visualisation tools. This problem will only
become more severe as we reach exascale computing. A plausible, attractive solution
involves processing data at simulation time (called in situ visualisation) to reduce the
data that must be transferred over networks and stored and to prepare the data for
more cost-effective post-processing visualisation. In this approach the strategy
is to have both the simulation and visualisation calculations run on the same parallel
supercomputer so that the data can be shared. Such simulation-time co-processing
can render images directly or extract features which are much smaller than the full
raw data to store for later examination. So, reducing both the data storage and transfer cost early in the data analysis pipeline optimizes the overall scientific discovery
process. In situ visualisation presents many new challenges to both simulation and
visualisation scientists. Before realising this approach, researchers must answer several questions, e.g. about the way simulation and visualisation calculations share the
same processor, memory space, and domain decomposition, what fraction of the supercomputer time should we devote to in situ data processing and visualisation and
if existing commercial and open source visualisation software tools can be directly
extended to support in situ visualisation at extreme scales.
\par
Mobile 3D maps portray the real environment as a virtual one, similar to their
desktop counterparts, but they run or should run in mobile devices. In mobile devices,
the computational power, memory, storage, and networking capabilities are increasing. They are being equipped with graphics hardware. For the first time, it might be
possible to portray the environment with direct one-to-one mapping as 3D, real-time
rendered mobile virtual environments. With wireless networking capabilities and G-
PS tracking, these environments could even be populated with real-world entities,
such as people and vehicles. Developing mobile 3D maps is no longer hindered by the
lack of 3D programming interfaces: the underlying rasterizer OpenGL ES is well sup-
ported on a range of devices, as is the Java-based scene graph renderer JSR-184. In
addition, the VRML viewing library Pocket Cortona is available for MobileWindows,
as is Direct3D Mobile. Photorealistic 3D cities from C3 Technologies, the leading
provider of 3D mapping solutions, are now available on Ovi Maps, Nokias free map-
ping and location service.
The Google Maps Javascript API lets you embed Google Maps in your own web
pages. Version 3 of this API is especially designed to be faster and more applicable to mobile
devices, as well as traditional desktop browser applications. The API provides a number of
utilities for manipulating maps and
adding content to the map through a variety of services, allowing you to create robust maps
applications on your website.

\subsection{Health}
The adoption of Information and Communication Technologies for sustainable healthcare is a main pillar of the European Digital Agenda for 2020, intends to develop and make available by 2015 with secure access to their online medical health data not just at home and travelling anywhere in the EU.
The development of new protocols for the better exploitation of the extremely sophisticated ICT technologies can definitely improve the quality of care available to Europe's patients and medical specialists.
\par Conventional areas of exploitation are telemedicine services, such as online medical consultation and personalized devices to check health conditions on real time in remote sites.
Other interventions are related to fundamental areas of the Information and Communication Technologies such as the epidemic spreading modelling \cite{N7}.
\par
Availability and quality of data, development of suitable algorithms and  models is becoming a must for doctors and technicians.
``Good health is ever more a question of good data, with health services in the middle of an information revolution, vital information will become more understandable - not necessarily more complex - as the data becomes accessible. ``\emph{Open Health Data}'' can bring transparency, innovation and efficiency into the public health system (see Open Data in Box 1).

\subsection{Critical Infrastructure Security}
In order to achieve the FuturICT vision, a wide range of S\&T challenges needs to be mastered
for providing the required capabilities, methods, and technologies. Key challenges include (i) advancing multi-sector critical infrastructure modelling, simulation and analysis
(ii) advancing the integration of external factors (human-made and natural
threats) and malicious attack  (to avoid a widened gap between our ability  to design and operate those systems and our capabilities to understand their complex behavior, starting from a recent state of the art assessment \cite{Kroeger,Hammerli,Brunner,WEF1,Weinberger,Simonsen,Peters,HelbingAK,Klein,Kuhl,Luiijf,Nieuwenhuijs,Pederson,Rinaldi}. Particular breakthroughs include:
\begin{itemize}
\item Integrated system analysis and modelling methods able to incorporate dynamic
physical models with human behaviour models, cyber layer models, and models
of external threats like natural hazards.
\item Improved scenario-based modelling and simulation methods that will make it easier to identify static and dynamic dependencies, to predict the behaviour of net-
worked systems.
\item  Middleware for semantic interoperability of distributed simulation and analysis
systems.
\item Advanced threat evaluation and vulnerability analysis tools.
\item  Improved  risk assessment tools that are able to regard organisational
models and human behaviour.
\item Advanced consequence and impact analyses tools for understanding the severity of
potential attacks, and their variability across all systems and critical
infrastructure sectors for different threat scenarios.
\item The capability to anticipate potential innovative uses and manipulation of new
technologies that can be turned against critical infrastructure, for example abuse
or attack of IP-addressable devices.
\item The capability to study and understand the vulnerabilities of technical critical
infrastructures systems, economical systems, and social organisations.
\item The capability to study and understand sociological and behavioural phenomena
that may lead people to commit acts of cyber-crime.
\end{itemize}

\section{Existing Initiatives and Projects}

The idea of the smart city has
been discussed since area wide networks were first mooted in the 1980s. Singapore
for example, badged itself in the late 1980s as the `\emph{Intelligent Island}' and put in place
several schemes relating to transportation (road pricing, driverless subway trains)
using the ICT that was available at the time. In fact it was James Martin who first
coined the term \emph{Wired Society} in the late 1970's that led to the idea of \emph{Wired Cities} in
the 1980s. Since then there have been many initiatives, first with public agencies
making plans for fibre optic cabling which have largely now been taken up and are
provided in a routine sense by telecoms, new online data systems being input using
web based access, for example for traffic control, then more routinised provision of
urban services as part of the idea of virtual cities, and then the rapid move of all kinds
of online access about the city, particularly with respect to maps using every kind
of information device.
\par
\emph{Smart Cities} involve making cities more efficient in several
different ways and thus the term is often used differently in different contexts. There
is now a strong business ethic to the idea of the smart city with ICT being considered
as the glue that holds the key to ways in which greater competitive benefits can be
realised through much greater integration and control of infrastructures that were
previously disconnected from one another. Caragliu et al. \cite{Caragliu} have produced a useful
survey paper on the proliferation of the smart city idea, identifying five key issues
that define the idea. These are:
\begin{itemize}
\item networked infrastructure to improve economic and
political efficiency, \item business-led urban development,\item the achievement of social
inclusion of various urban residents in public services, \item a focus on the crucial role
of high-tech and creative industries in long-run urban growth, \item a focus on learning,
adaptation and innovation, \item sustainability as a major strategic component of a
city is smartness.
\end{itemize}
\par
There are five kinds of major projects currently associated with smart cities:
\begin{enumerate}
  \item Key initiatives being mounted by the world's major IT companies as they see
their market for future software being to merge with governance advices which
are accessible by the public at large.
\par
 Oracle is active in promoting the idea of
integrated network services, Microsoft see their role as supporting new initiatives such
as the city in Portugal being built by \emph{Living Planet-IT} and Ericsson proposes that
the delivery of the smart city is best explored in virgin connects such as those being developed in newly industrialising counties  and the integration of what have hitherto been separate services. In particular, the  \emph{IBM Smarter Planet} initiative is key in that they are developing a variety
of generic software which will access, mine and monitor online data that is being
used to some purpose, particularly in areas of water quality, health and pollution and traffic. Cisco Systems have developed a vision of the Connected Community which is strongly
associated with the wired cities and the provision of smart services.
Indeed, all the major players in ICT have their visions for the smart city agenda
which largely involves the provision of new software and services built on their
own hardware that are city-wide
\item Major initiatives in the EU in smart governance which involve developing a) decision support tools for dealing with online data about cities and for enabling
wide participation in generating scenarios for the smart city of the future, and b)
demonstrator projects associated with many local government in the EU which
are being used to demonstrate the value added of online citizen services. Into this
nexus comes a major thrust for citizen science which often focussed at the city
level.
\item Major initiatives by telecoms companies dealing with the provision of new forms of
network and their coordination which build on ideas about the networked future.
Eriksson for example, have a strong interest in these developments.
\item New forms of social media being networks with a strong localism agenda and this
is leading to all kinds of new social networks that need to be integrated with
other more conventional networks involving flows of information and materials
using telecoms and transportation systems.
\end{enumerate}
\par
At the moment there are no ready simulation tools available which have the scale and complexity to simulate the linkages across the supply chains and between
industries and between the supply chains and the consumers. The on-going work
on MATSim has demonstrated the possibilites, with productive implementations for
Switzerland, Singapore, LA basin, Netherlands, Guateng Province involving millions
of agents and navigation networks travelling between millions of destinations. This
model framework is able to capture the linkages within the daily life of the residents of
such regions and countries.
There is additional work, in particular, on the resilience of
scheduled airline services.
\par
The complexity science driven work on gridlock is currently too abstract to provide engineering
or policy guidance, but is supplemented by policy oriented spatial regression based
work. The ENCOURAGED (Energy corridor optimisation for European markets of
gas, electricity and hydrogen) project has been launched in 2005 to identify and
assess the economically optimal energy corridors between European Union (EU) and
neighboring countries. The DESERTEC concept aims at promoting the generation
of electricity in Northern Africa and the Middle East by using solar power plants
and wind parks and the transmission of this electricity to Europe. The MEDGRID
is promoting new high capacity electricity links around the Mediterranean \cite{N17}.
\par
The 7$^{th}$ Framework Programme of the European Commission has generated some 36 European Technology Platforms (ETP) and Joint Technology Initiative (JTI) that are relevant to the Exploratory of Technology. Establishing links to all these
JTIs would enable early acquisition of information on planned technology developments.
Links should be established to relevant projects in the context of the
European Parliament Technology Assessment (EPTA). Experts conducting Technology Assessments for EPTA (like [EPTA-EM, EPTA-LT]) should be attracted for
contributing expertise to the Technology Exploratory.
\bigskip
\begin{mdframed}[backgroundcolor=shadecolor]
\small
\begin{center}
\textbf{BOX 2:
Relevant Initiatives and Projects}
\end{center}
\begin{itemize}
\item EU FP7 Research Infrastructures Design Study DIESIS, \url{http://www.diesis-project.
eu/}.
\item EU FP7 Security STReP EMILI, \url{http://www.emili-project.eu/}.
\item European Network and Information Security Agency, \url{http://www.enisa.europa.eu/}.
\item European Public-Private Partnership for resilience, \url{http://ec.europa.eu/
information_society/policy/nis/strategy/activities/ciip/ep3r/index_en.htm}
\item European Programme for Critical Infrastructure Protection, \url{http://europa.eu/
legislation_summaries/justice_freedom_security/fight_against_terrorism/
l33260_en.htm}
\item European Parliamentary Technology Assessment, \url{http://www.eptanetwork.org/}
\item{EPTA} Study Electric mobility concepts and their significance for economy, society, environment,
\url{http://www.tekno.dk/EPTA/projects.php?pid=736}
\item Study Localisation technologies, \url{http://www.ta-swiss.ch/en/projects/
information-society/localisation-technologies/}
\item European Commission (2009): Council Regulation (EC) No 723/2009 of 25 June 2009
on the Community legal framework for a European Research Infrastructure Consortium
(ERIC), Official Journal of the European Union L206/1, 8.8.2009.
\item European Commission (2009): Practical Guide for the Community legal framework for
a European Research Infrastructure Consortium (ERIC), Version 1 27/11/09
\item Individual European Technology Platforms, \url{http://cordis.europa.eu/
technology-platforms/individual_en.html}
\item European Commission (2008): COUNCIL DIRECTIVE 2008/114/EC of 8 December
2008 on the identification and designation of European critical infrastructures and the
assessment of the need to improve their protection, Brussels, December 8, 2008. European programme for the establishment of a European capacity for Earth Observation, (GMES)
\url{http://www.gmes.info/index.php}
\item  Individual Joint Technology Initiatives, \url{http://cordis.europa.eu/fp7/jtis/}
\item EU COST Action IC0806 IntelliCIS, \url{http://www.intellicis.eu/}
\item  The RECIPE Project: Critical Infrastructure Protection (CIP) Policies In Europe
%\item  US Department of Homeland Security: National Infrastructure Simulation and Analysis
%Center, USA, \url{http://www.sandia.gov/nisac/}
\item  OECD (2008): Protection of Critical Infrastructure and the role of investment policies
relating to national security, May 2008.
\end{itemize}
\end{mdframed}
\bigskip
\normalsize
\bigskip
\par
In 2000, the National Infrastructure Simulation and Analysis Center (NISAC) was
founded in the USA as part of a cooperation between the Sandia and Los Alamos
National Laboratories. In 2003, NISAC was integrated into the Department of Homeland Security. NISAC has a large inventory of models and simulation and analysis
tools for evaluating failures of CI, for fast assessment of impact and damage of natural
hazards, and more. NISAC is of a comparable scale as the Technology Exploratory.
In January 2011, a five-years project on Long term dynamics of interdependent infrastructure systems started in the UK (ITRC). The aim of the project is to develop and
demonstrate a new generation of system simulation models and tools to inform analysis, planning and design of National Infrastructures. The project does not consider
social factors and developments and is of a much smaller scale than the Technology
Exploratory.
\par
The EU has funded several dozen R\&D projects on security and Critical
Infrastructures, both in its Research Framework Programmes and in funding schemes
of DG HOME (formerly JLS) and COST, for instance (IRRIIS, EMILI, RECIPE,
IntelliCIS, OpenMI). All of these projects are or were on a smaller scale than the
Technology Exploratory. So far, there is no sustainable facility in Europe that could
be compared to the Technology Exploratory. EU FP7 Capacities has funded the Design Study DIESIS that proposes a trans-national European facility for collaborative
R\&D for protecting Critical Infrastructures (called EISAC). Many of the concepts of
the DIESIS Design Study could be adopted for the Technology Exploratory. Within
the European programme for the establishment of a European capacity for Earth Observation (Global Monitoring for Environment and Security) (GMES), a number of
projects have been set up for realising pre-operational services. It needs to be checked
if and how these services could be employed for the Exploratories.
\section{Matching of Technology Exploratory with the other FuturICT Activities}
The ``Planetary Nervous System" might be able to produce data that are relevant for
simulations within the ``Living Earth Simulators" and ultimately used for decision
and strategy making within the ``Global Participatory Platform". The basic technologies described above are example of the key components for these platforms. The
``Planetary Nervous System'' will operate over a network of networks made of wireless
and wired computers, phone and sensors and thus will rely on the availability of the
most advanced technologies for those components. The same need applies to the ``Living Earth Simulator" and to the ``Global Participatory Platform" which will rely on
the availability of last-generation computational and visualization infrastructures for
their operability.
The availability of new technology is not enough to ensure social well-being. Transforming new technology into innovation can only be achieved upon availability of economic and political commitment and social preparedness. The Technology Exploratory will benefit of the Innovation Accelerator to improve effectiveness and efficiency
of existing policies in promoting innovation (for example renewable energy support,
economic and political aspects of innovation policies, electricity and transportation
market price behaviour, efficient pricing and investment, regulatory behaviour). Results of the ``Innovation Accelerator" could be used for the innovation scouting part of the Technology Exploratory.
Assessments and forecasts of consequences of failures of Critical Infrastructures
could provide input to the Exploratory of Economics. Economic consequences of
technological development that influences Critical Infrastructures (Smart Grids, permeation of SCADA systems with Internet, etc.) could be provided by the Exploratory
of Economics for use in the Technology Exploratory.
The Society Exploratory could provide data on society's dependencies on Critical
Infrastructures, how society uses, adopts technological changes and reacts to failing
of Critical Infrastructures. Such data will contribute to improved risk assessment and
management. Simulations of natural hazards and forecasts of damages provided by
the Exploratory on Environment could and should be employed for scenarios of Critical Infrastructures risk assessment in the Technology Exploratory. for the following
reasons. Some studies state that natural causes are most common reasons for Critical Infrastructures failures. Although TNO's Critical Infrastructure outage database
shows this is less the case in the EU as compared to the rest of the world, natural
causes are a main risk factor. Due to the climate change, the high number of natural
disasters is likely to sustain or even increase. A 2008 paper from the
High Representative and the European Commission to the European Council on climate change and international security (HREC) recommends, among other measures,
to intensify EU capacities for research, analysis, monitoring and early warning. The
latter two shall include threats to Critical Infrastructure and economic assets.
Need to pool resources to address such challenge Due to the holistic nature of
the system of systems level approach, the Security/Critical Infrastructures sector of
the Exploratory of Technology requires the cooperation of computer scientists, ICT
experts in modelling, simulation and analysis, experts from various Critical Infrastructures sectors (energy, water, finance, telecommunication, etc.), experts in (cyber)
security, social scientists, and more.
Critical Infrastructures, such as water supply, energy and telecommunication systems, are vital for society. Due to technical, political, and economical changes, Critical
Infrastructures evolve and are becoming increasingly complex and intertwined. New
technologies, such as the World Wide Web, can be adopted at breath-taking speed
and may lead to significant changes of society and economy. In order to ensure the
continuity of the vital services of these infrastructures, it is essential to understand
in detail their complex behaviour and current and future threats endangering their
functioning. Because of the dependencies among different Critical Infrastructures and
their cross-organisational, cross-sector and cross-border nature, no organisation, sector or country can reach this understanding on its own. Improved understanding
and development of methods for increasing the resilience of Critical Infrastructures requires the cooperation of researchers, stakeholders, and authorities in order to create a critical mass of expertise.

\begin{figure}
\center
% Use the relevant command for your figure-insertion program
% to insert the figure file.
% For example, with the option graphics use
\resizebox{1.0\columnwidth}{!}{%
\includegraphics{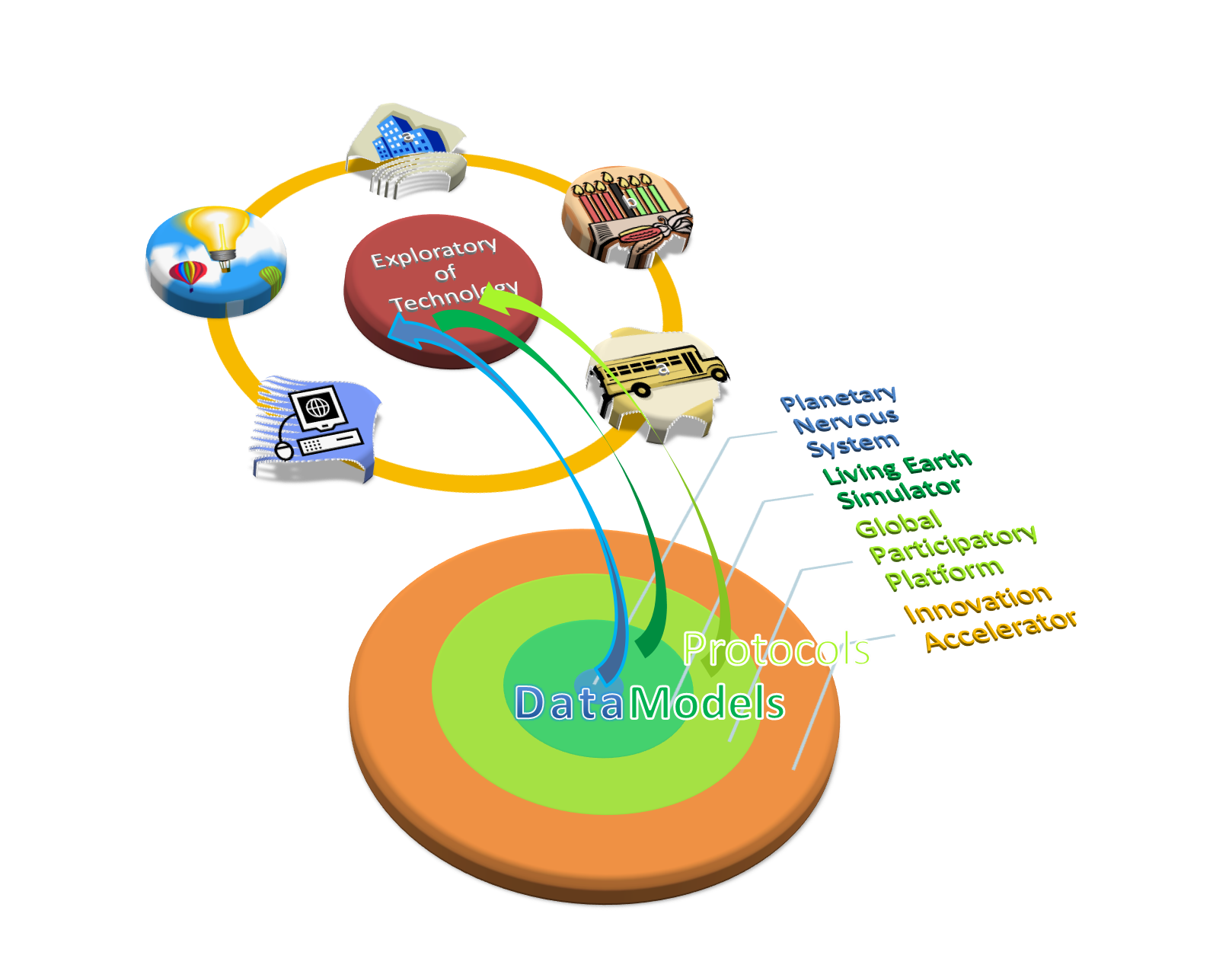}}
\caption{Scheme of the interrelationship between the Exploratory of Technology and other
FuturICT components.}
\label{fig:6}       % Give a unique label
\end{figure}

\section{Acknowledgements}
We are grateful to the anonymous reviewers for many insightful comments.
The publication of this work was partially supported by the European
Union's Seventh Framework Programme (FP7/2007-2013) under grant agreement no.
284709, a Coordination and Support Action in the Information
and Communication Technologies activity area (`FuturICT' FET Flagship Pilot Project).

\end{document}